\documentstyle[preprint,prc,aps,epsfig]{revtex}
\tightenlines
\begin{document}
\title{An Analysis of the $^{12}$C+$^{24}$Mg Reaction Using
A New Coupling Potential}
\author{I. Boztosun \footnote[1]{Permanent address
: Department of Physics, Erciyes University, Kayseri 38039 Turkey}
\footnote[2]{Present address: Computational Mathematics Group,
School of Computer Science and Mathematics, University of
Portsmouth, Portsmouth PO1 2EG UK} and W.D.M. Rae}
\address{Department of Nuclear Physics, University of Oxford,
Keble Road, Oxford OX1 3RH UK}
\date{\today}
\maketitle
\begin{abstract}
We introduce a new coupling potential to explain the experimental
data for the $^{12}$C+$^{24}$Mg system at numerous energies in the
laboratory system from 16.0~MeV to 24.0 MeV. This new
coupled-channels based approach involves replacing the usual first
derivative coupling potential by a new, second-derivative coupling
potential. This paper first shows and discusses the limitation of
the standard coupled-channels theory in the case where one of the
nuclei in the reaction is strongly deformed. Then, this new
approach is shown to improve consistently the agreement with the
experimental data and has made major improvement on all the
previous coupled-channels calculations for this system.
\end{abstract}
\section{Introduction}
The elastic and inelastic scattering of the light heavy-ion
reactions such as $^{16}$O+$^{28}$Si, $^{12}$C+$^{24}$Mg and
$^{12}$C+$^{12}$C have been extensively investigated over the last
40 years and a large body of experimental data has been
accumulated from the systematic studies of these reactions (see
\cite{Bra82,Sci97,Sto79} and references therein). A variety of
theoretical accounts, based on dynamical models or purely
phenomenological treatments, have been proposed to explain the
experimental data \cite{Bra82,Bra97,Kob84}. However, there appears
no unique model that explains consistently the elastic and
inelastic scattering data over wide energy ranges without applying
any {\it ad-hoc} procedures.

Consequently, the following problems continue to exist for the
light heavy-ion reactions \cite{Boz1,Bozth}: $(1)$ explanation of
anomalous large angle scattering data; $(2)$ reproduction of the
oscillatory structure near the Coulomb barrier; $(3)$ the
out-of-phase problem between theoretical predictions and
experimental data; $(4)$ the deformation parameters ($\beta$
values): previous calculations require $\beta$ values that are at
variance with the empirical values and are physically
unjustifiable.

The elastic and inelastic scattering data of the
$^{12}$C+$^{24}$Mg system have been studied extensively and some
of the above-mentioned problems could not be accounted for
\cite{Sci97,Car76,Car78,Fil89}. The most extensive study for this
system was carried out by Sciani {\it et al} \cite{Sci97} who used
$Q$-dependent potentials whose parameters had different values for
the incoming and outgoing channels in the coupled-channels
calculations. Without $Q$-dependent potentials, they observed that
the theoretical calculations and the experimental data were
completely out-of-phase and could not reproduce the experimental
data. However, they overcame this problem by introducing these
$Q$-dependent potentials. Nevertheless, not only were the
parameters changing from energy to energy in an arbitrary way, but
they also had to change the $\beta$ value in order to optimize the
fits.

It has been the practice to increase or decrease artificially the
$\beta$ value to obtain the magnitude of the 2$^{+}$ state data
correctly in the standard coupled-channels calculations, without
giving the physical justification other than stating that it is
required to fit the
data~\cite{Sci97,Car76,Car78,Bra77,Dud78,Bra85}.

The out-of-phase between the theoretical predictions and the
experimental data for the ground and 2$^{+}$ states has also been
observed and without optimizing the $\beta$ value, it has been
impossible to obtain a simultaneous fit to the elastic and
inelastic scattering data \cite{Sci97,Boz99,Bra77,Dud78,Bra85}.

Therefore, building on a previous paper \cite{Boz1}, which was
outstandingly successful in explaining the experimental data for
the $^{12}$C+$^{12}$C system, we investigate the
$^{12}$C+$^{24}$Mg reaction which has been intensively
investigated experimentally at energies near the Coulomb
barrier~\cite{Sci97,Car76,Car78,Fil89}. The main feature of the
experimental data is a strong oscillatory structure which can not
be explained in a wide energy range within the coupled-channels
and DWBA methods if any {\it ad-hoc} procedures are not applied.
In this paper, our aim is to explain the elastic and inelastic
scattering data with empirical $\beta$ value.

In the next section, we first introduce the standard
coupled-channels model and show the results of these analyzes in
section \ref{stanres} from E$_{Lab}$=16.0 MeV to 24.0 MeV. Then,
in section \ref{newcc}, we introduce a new coupling potential to
analyze the experimental data in the same energy range and show
the results of these new coupled-channels calculations. Finally,
section \ref{conc} is devoted to our summary and conclusion.
\section{The Standard Coupled-Channels Calculations}
\label{stan}
\begin{figure}
\epsfxsize 8.5cm \centerline{\epsfbox{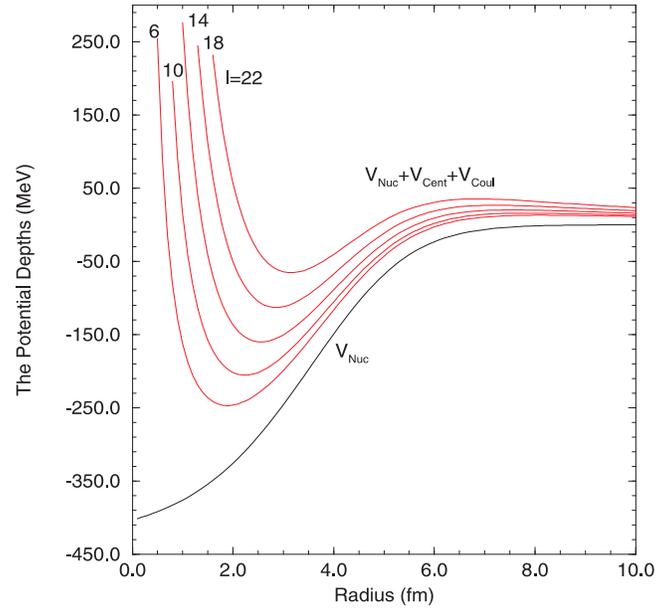}} \vskip+0.5cm
\caption{The interaction potential between $^{12}$C and $^{24}$Mg
is plotted for various values of the orbital angular momentum
quantum number, $l$} \label{real}
\end{figure}

The interaction between $^{12}$C and $^{24}$Mg nuclei is described
by a deformed optical potential. As shown in figure \ref{real},
the real potential is chosen as the square of a Woods-Saxon shape:
\begin{equation}
V_{N}(r) = \frac{-V_{0}}{(1+exp(r-R)/a)^{2}} \label{realpot}
\end{equation}
and the parameters are fixed as a function of energy to reproduce
the experimental data over the whole energy range. The numerical
values are shown in table \ref{realparam}.

The sum of the nuclear, Coulomb and the centrifugal potentials is
also shown in the same figure for various values of the orbital
angular momentum quantum number, $l$. The superposition of the
attractive and repulsive potentials results in the formation of a
potential pocket, which the width and depth of the pocket depend
on the orbital angular momentum. This pocket is very important for
the interference of the barrier and internal waves, which produces
the pronounced structure in the cross-section. The effect of this
pocket can be understood in terms of the interference between the
internal and barrier waves that correspond to a decomposition of
the scattering amplitude into two components, the inner and
external waves~\cite{Lee78,Bri77}.

The imaginary potential has the standard Woods-Saxon volume shape
as in equation (\ref{imagpot}) and the depth increases linearly
with energy as in equation (\ref{cmgimag}).
\begin{equation}
W(r)=-\frac{W_{0}}{(1+exp((r-R)/a))} \label{imagpot}
\end{equation}
\begin{equation}
W = 4.375E_{Lab}-67.0 \label{cmgimag}
\end{equation}
The other parameters of the real and imaginary potentials are
fixed as a function of energy and are not changed in the present
calculations as shown in table \ref{realparam}.

It is assumed that the target nucleus $^{24}$Mg has a static
quadrupole deformation and this assumption is taken into account
by deforming the real potential in the following way.
\begin{equation}
R(\theta,\phi)=r_{0}A_{P}^{1/3}+r_{0}A_{T}^{1/3}[1+\beta_{2}
Y_{20}(\theta,\phi)]
\end{equation}
where $P$ and $T$ refer to projectile and target nuclei
respectively and $\beta_{2}$ is the deformation parameter of
$^{24}$Mg. In the present calculations, only target nucleus,
$^{24}$Mg, is deformed although it is well known that the
projectile $^{12}$C is also a strongly deformed nucleus. However,
when it is deformed, the number of channels increases and it makes
the computational processing time insurmountable.

In our  coupled-channels calculations, we shall use the exact
value of $\beta_{2}$, derived from the deformation length,
$\delta$. The invariant parameter in the coupled-channels
formalism is in fact the deformation length, $\delta$=$\beta$R, or
its value derived from the reduced electromagnetic transition
probability, B(E2), rather than $\beta$ itself.

The actual value of B(E2) is 430 e$^{2}$fm$^{4}$ \cite{Chr73} and
that of $\delta$ is between 1.48 fm \cite{Tho77} and 1.50 fm
\cite{Car76} for the target nucleus $^{24}$Mg. We use
$\delta$=1.50 fm ($\beta$=0.52) in our coupled-channels
calculations. For the Coulomb deformation, we assume
$\beta_{2}^{C}$=$\beta_{2}^{N}$ \cite{Sat83}.

In the present calculations, the first two excited states of the
target nucleus $^{24}$Mg, {\it i.e.} 2$^{+}$ (1.47 MeV) and
4$^{+}$ (4.12 MeV), are included and the 0$^{+}$-2$^{+}$-4$^{+}$
coupling scheme was employed. The reorientation effects for
2$^{+}$ and 4$^{+}$ excited states are also included. An
extensively modified version of the code CHUCK \cite{Kunz} has
been used in all the calculations.
\section{Results}
\label{stanres}
\begin{figure}
\epsfxsize 10.5cm \centerline{\epsfbox{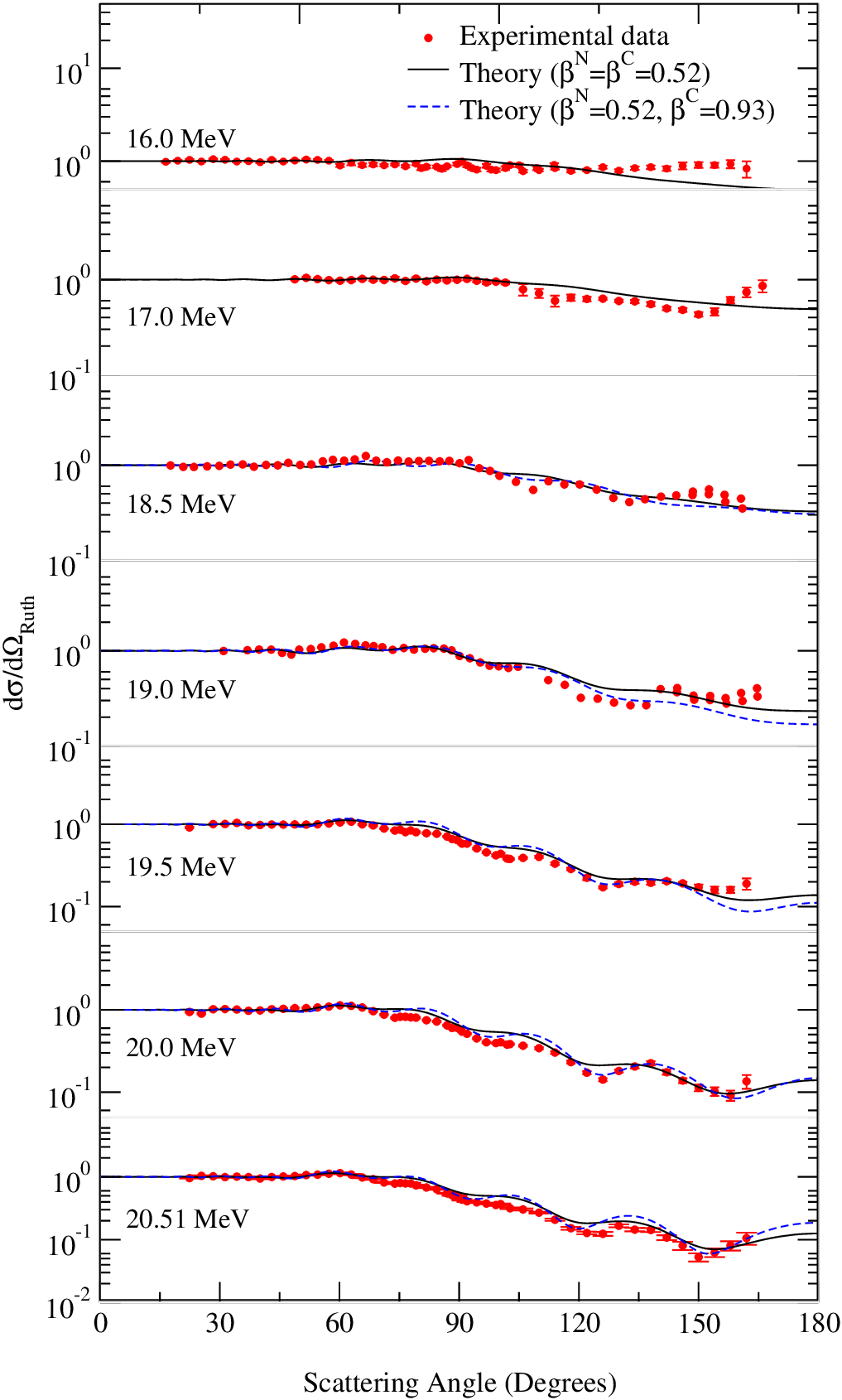}}
\vskip+0.5cm \caption{Ground state results of the standard
coupled-channels calculations with
$\beta_{2}^{N}$=$\beta_{2}^{C}$=0.52 (solid lines) and with
$\beta_{2}^{N}$=$\beta_{2}^{C}$=0.93 (dashed lines).}
\label{ws2ground1}
\end{figure}

\begin{figure}
\epsfxsize 10.5cm \centerline{\epsfbox{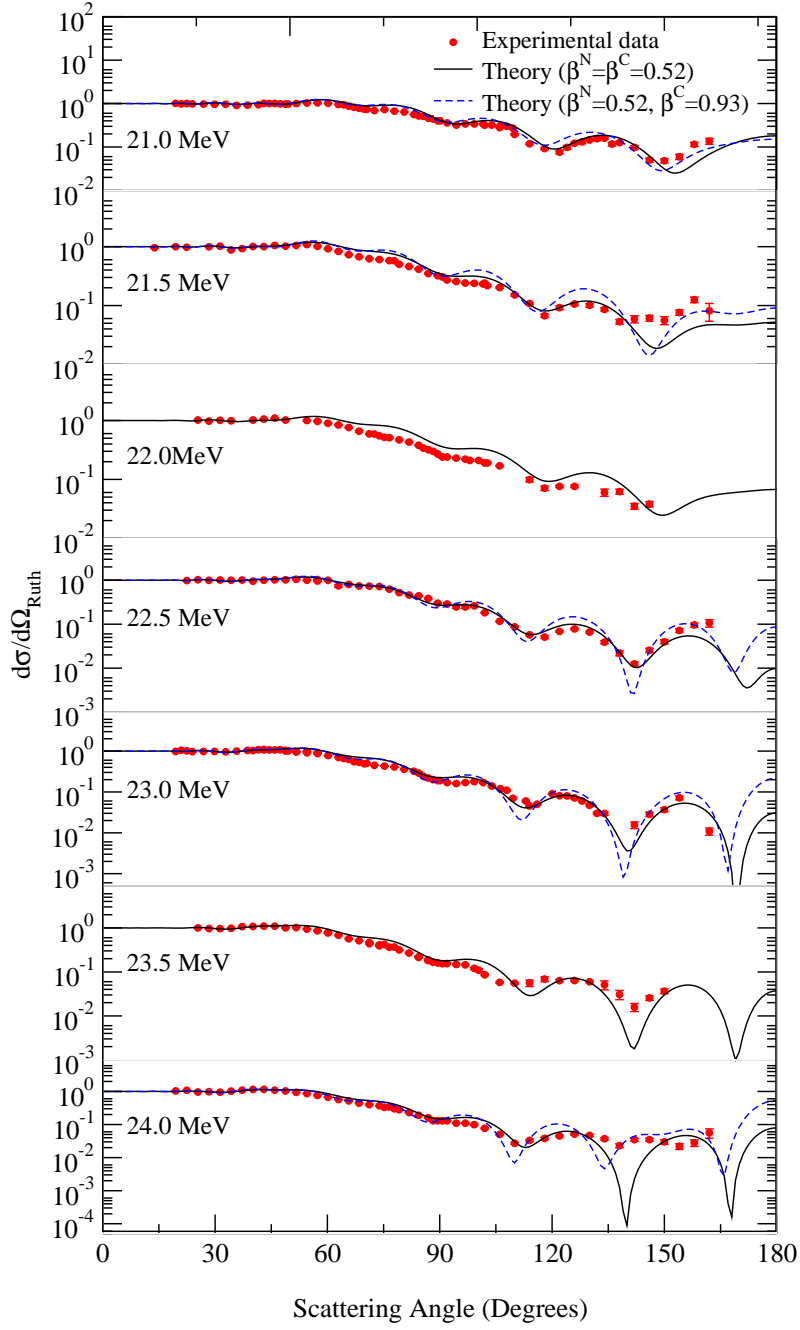}}
\vskip+0.5cm \caption{Ground state results of the standard
coupled-channels calculations with
$\beta_{2}^{N}$=$\beta_{2}^{C}$=0.52 (solid lines) and with
$\beta_{2}^{N}$=$\beta_{2}^{C}$=0.93 (dashed lines) ({\it
continued from figure \ref{ws2ground1}}).} \label{ws2ground2}
\end{figure}

\begin{figure}
\epsfxsize 8.5cm \centerline{\epsfbox{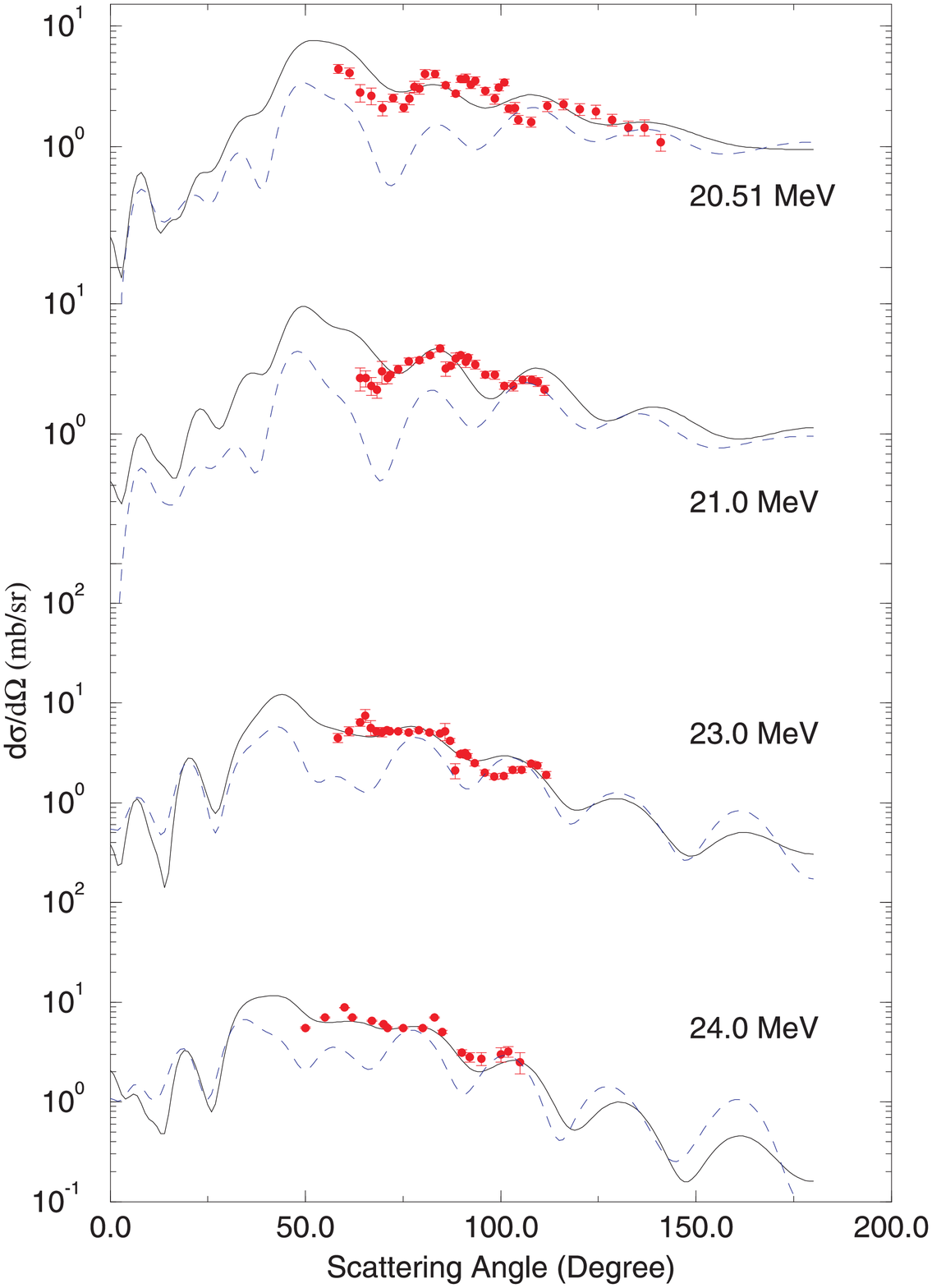}}
\vskip+0.25cm \caption{The 2$^{+}$ state results of the standard
coupled-channels calculations with
$\beta_{2}^{N}$=$\beta_{2}^{C}$=0.52 (dashed lines) and with
$\beta_{2}^{N}$=$\beta_{2}^{C}$=0.93 (solid lines).}
\label{ws2single}
\end{figure}

Using the standard coupled-channels model, some of the results
obtained using the empirical beta values for the nuclear and
Coulomb deformations ($\beta_{2}^{N}$=$\beta_{2}^{C}$=0.52) are
shown in figures \ref{ws2ground1} and \ref{ws2ground2} for the
ground state and in figure \ref{ws2single} for the first excited
state with dashed lines respectively. Although we obtained a good
agreement with the experimental data for the ground state, it has
not been possible to get the magnitude of the first excited state
(2$^{+}$) correctly. The magnitudes of the 2$^{+}$ predictions are
smaller than the measured experimental data and the minima and
maxima observed in the experimental data are not reproduced
correctly.

This has been a recurring problem in earlier theoretical
calculations, where numerous arbitrary values of the deformation
parameter had to be
used~\cite{Sci97,Car76,Car78,Bra77,Dud78,Bra85}. Varying the
parameters and changing the shape of the real and imaginary
potentials do not provide a global fit to the experimental data
for both ground and 2$^{+}$ states. However, it is clear from
these results that the problem is in the forward angle region,
where two factors are very important \cite{Sat83,Tam65}. The first
one is the number of partial waves used in the calculations. Since
the calculated cross-section depends on the orbital angular
momentum number, $l$, the number of the partial waves used in the
calculations should be checked whether they all contribute. This
conception is examined in order to determine their effect on the
results and it is observed that the number of the partial waves do
not affect the results beyond a critical value.

The second factor, which is effective in this region, is the value
of the Coulomb deformation parameter, $\beta_{2}^{C}$. The
sensitivity of the calculations to the $\beta_{2}^{C}$ is checked
and the $\beta_{2}^{C}$ value required to fit to the data is found
to be $\beta_{2}^{C}$=0.93, which is larger than its actual value.

The results of these calculations using the exact $\beta_{2}^{N}$
and increased $\beta_{2}^{C}$ are shown in figures
\ref{ws2ground1} and \ref{ws2ground2} for the ground state and in
figure \ref{ws2single} for the first excited state with solid
lines. The agreement  is good for both the ground and the first
excited states over the whole energy range although the magnitude
and the phase oscillation problems persist at high energies for
the 2$^{+}$ results.
\section{New Coupling Potential}
\label{newcc}
In the analyzes of this reaction, our aim was to solve the
out-of-phase problem and to reproduce the experimental data with
empirical $\beta$ value. We succeeded in achieving the former one,
but failed to provide a solution to the latter.

\begin{figure}
\epsfxsize 8.0cm \centerline{\epsfbox{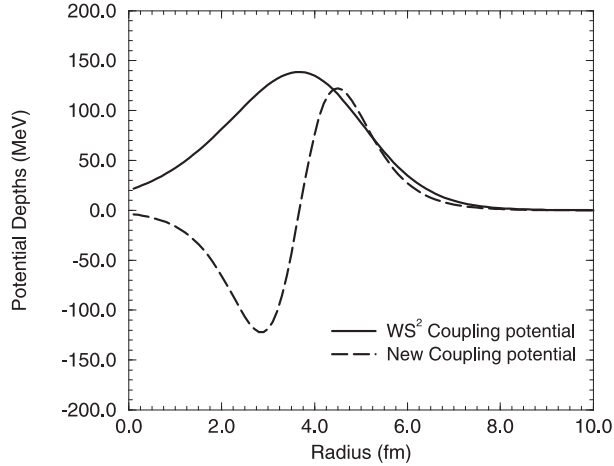}} \vskip+0.25cm
\caption{The comparison of the {\it standard} coupling potential,
which is the first derivative of the central potential, with our
{\it new} coupling potential, parameterized as the 2$^{nd}$
derivative of Woods-Saxon shape as in equation \ref{couppot}.}
\label{couplingpotcmg}
\end{figure}

Because of the limitations of the standard coupled-channels method
in the analyzes of this reaction, we use a new second-derivative
coupling potential which has successfully explained the
experimental data for the $^{12}$C+$^{12}$C reaction \cite{Boz1}.
The standard and new coupling potentials are compared in figure
\ref{couplingpotcmg} and the new coupling potential has the
following shape:

\begin{equation}
V_{C}(r) = \frac{-V_{C_{0}} \,\, e^{(r-R)/a}
(e^{(r-R)/a}-1)}{a^{2}\left[1+e^{(r-R)/a} \right]^{3}}
\label{couppot}
\end{equation}

where $V_{C_{0}}$=185.0 MeV, $R$=3.67 fm and $a$=0.62 fm.

One possible interpretation of such a second-derivative coupling
potential can be made if we express the total potential as a
function of the radii for different orientations of the two
colliding $^{12}$C and $^{24}$Mg nuclei. If $\theta_{P,T}$ are the
angles between the symmetry axes and the axis joining the centers
of the projectile and target, then the total potential, as an
approximation, can be expressed in the following way:
\begin{equation}
V(r)=V_{N}(r)+\beta_{2}R_{P}\frac{dV_{C}}{dR_{P}}Y_{20}(\theta_{P},
\phi_{P})+ \beta_{2}R_{T}\frac{dV_{C}}{dR_{T}}Y_{20}(\theta_{T},
\phi_{T}) \label{orientfor}
\end{equation}
where $V_{N}$ is the nuclear potential and $V_{C}$ is the new
second-derivative coupling potential. The difference between
equation 8 in ref. \cite{Boz1} and equation \ref{orientfor} is due
to the simultaneous mutual excitation of two nuclei. In ref.
\cite{Boz1}, we took in to account the simultaneous mutual
excitation of the projectile and target nuclei, therefore there is
an extra term to define it, whereas in equation \ref{orientfor} we
do not have mutual excitation term since we just include the
excitation of target nucleus.

\begin{figure}
\epsfxsize 9.0cm \centerline{\epsfbox{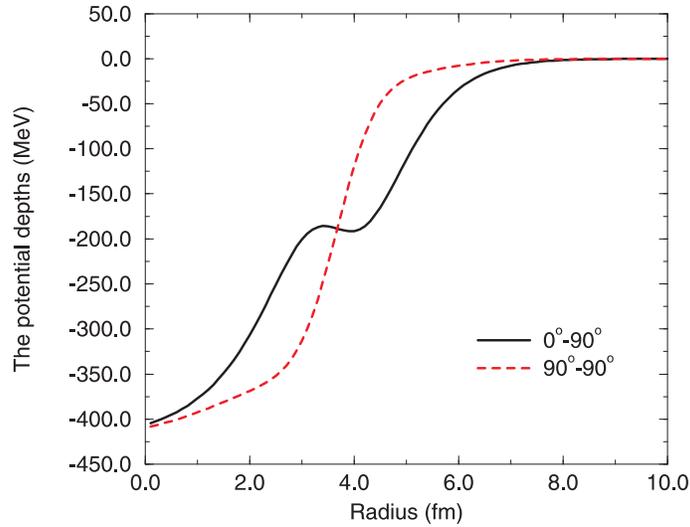}} \vskip+0.0cm
\caption{The orientation potentials of the $^{12}$C and $^{24}$Mg
nuclei at different angles.} \label{orient}
\end{figure}

The result for the  $^{12}$C+$^{24}$Mg system  is shown in figure
\ref{orient}. A second local minimum is observed in the
interaction potential for certain orientations. This feature has
not been taken into account in the standard coupled-channels
calculations. To investigate this minimum, we looked at the total
inverted potential, {\it i. e.} the dynamical polarization
potential (DPP) plus the bare potential, obtained by the inversion
of the S-Matrix \cite{Boz3}. Our analysis suggests that the new
coupling potential points to the presence of the super-deformed
configurations in the compound nucleus $^{36}$Ar, as it has been
speculated \cite{Rae88,Fes92}.
\section{Results}
\label{newres} The real and imaginary potentials in these new
calculations have the same shapes and parameters as in previous
calculations (see equation~\ref{realpot} and~\ref{imagpot}) and
the parameters of the new coupling potential are displayed in the
caption of figure \ref{couplingpotcmg}. We have analyzed the
experimental in the same energy range.

\begin{figure}
\epsfxsize 10.5cm \centerline{\epsfbox{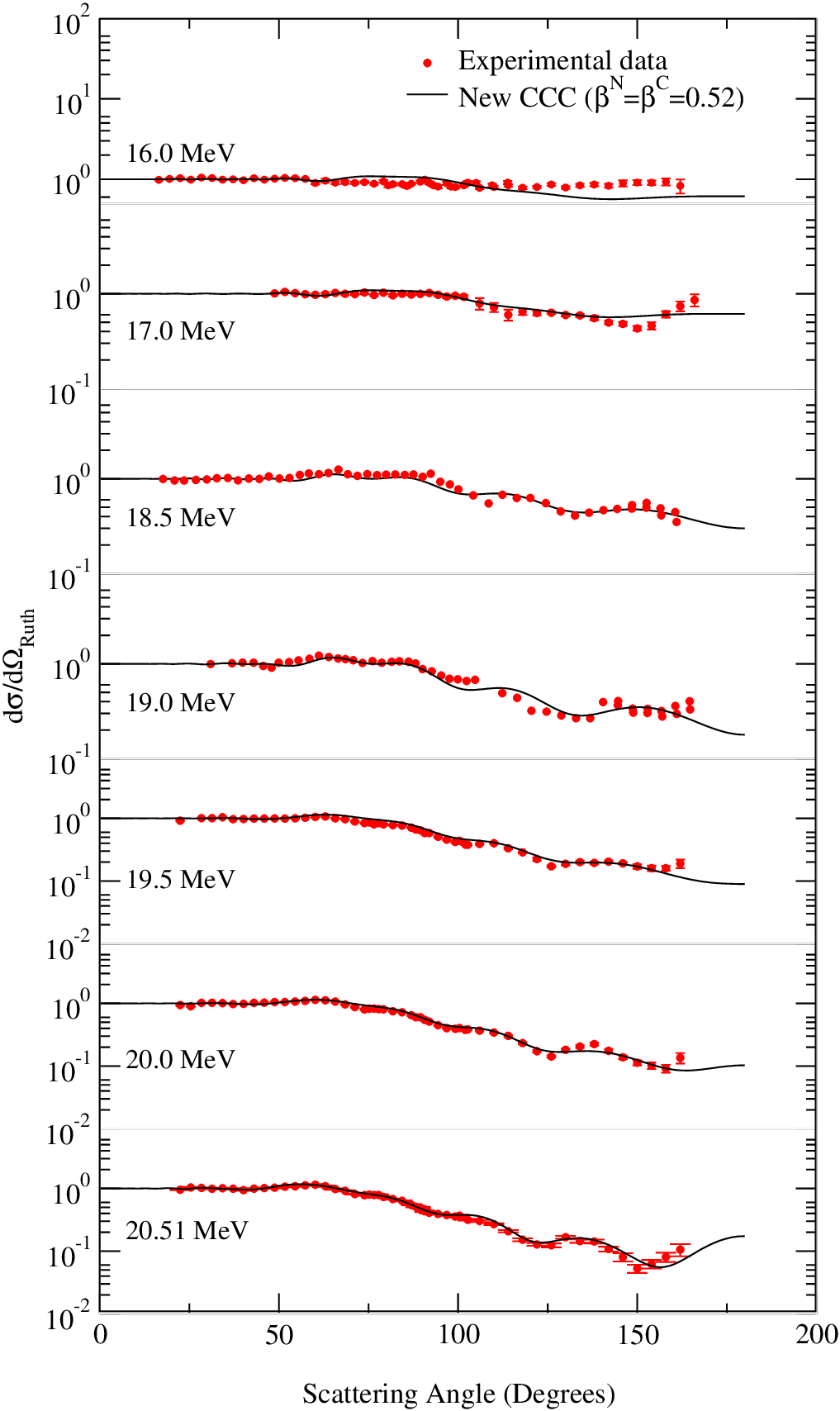}}
\vskip+0.5cm \caption{Ground state results of the new
coupled-channels calculations using the new coupling potential
with the exact $\beta$ value
($\beta_{2}^{C}$=$\beta_{2}^{N}$=0.52).} \label{2ndground1cmg}
\end{figure}

\begin{figure}
\epsfxsize 10.5cm \centerline{\epsfbox{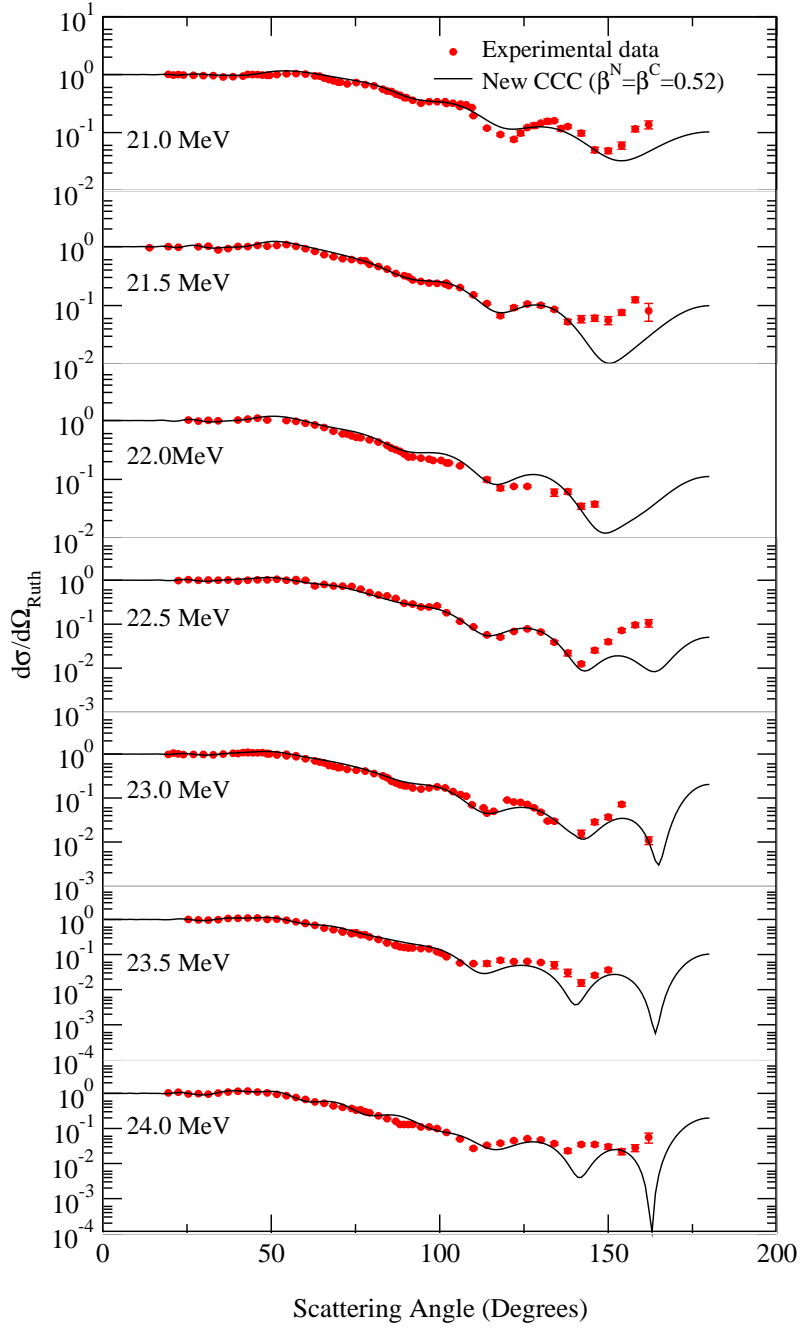}}
\vskip+0.5cm \caption{Ground state results of the new
coupled-channels calculations using the new coupling potential
with the exact $\beta$ value
($\beta_{2}^{C}$=$\beta_{2}^{N}$=0.52) ({\it continued from figure
\ref{2ndground1cmg}}).} \label{2ndcmgground2}
\end{figure}

\begin{figure}
\epsfxsize 8.5cm \centerline{\epsfbox{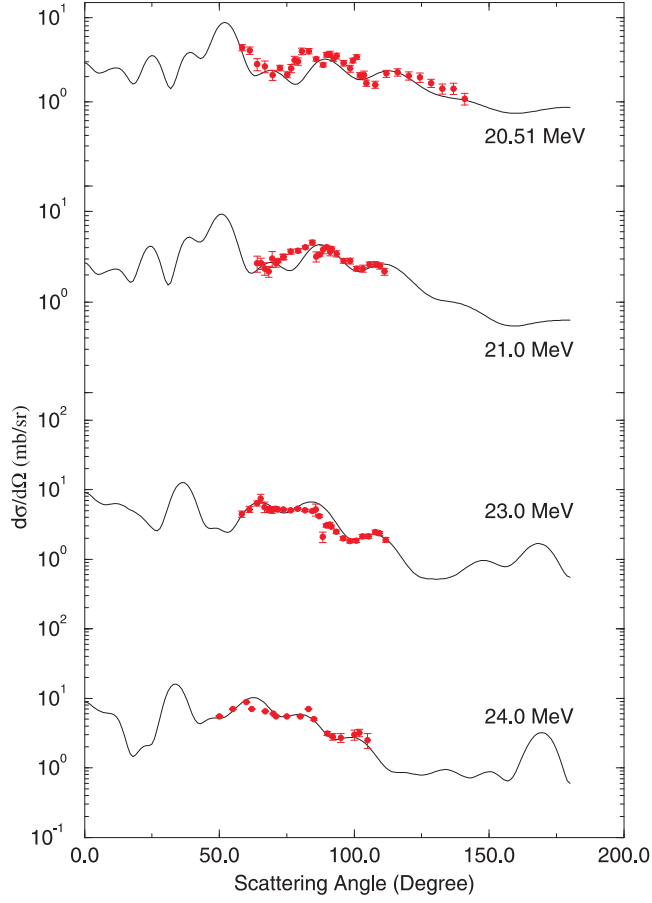}}
\vskip+0.20cm \caption{The 2$^{+}$ state results of the new
coupled-channels calculations obtained using the new coupling
potential with the exact $\beta$ value
($\beta_{2}^{C}$=$\beta_{2}^{N}$=0.52).} \label{2ndcmgsingle}
\end{figure}

It is clearly seen from figures \ref{2ndground1cmg} and
\ref{2ndcmgground2} for the ground state and figure
\ref{2ndcmgsingle} for the first excited state that the new
second-derivative coupling potential with the exact $\beta$ value
($\beta_{2}^{N}$=$\beta_{2}^{C}$=0.52) yields excellent agreement
with the experimental data over the whole energy range studied.
These figures show perfect fits with the experimental data; the
phases of the oscillations and magnitudes in the 2$^{+}$ state
data are well accounted for.

The comparison of the $\chi^{2}$ values in table \ref{chi1}
indicates that this new coupling potential has not only solved the
out of phase problem and reproduced the experimental data with
empirical $\beta$ value, but also improved the quality of the
fits.
\section{Summary}
\label{conc}
We have shown a consistent description of the elastic and
inelastic scattering of the $^{12}$C+$^{24}$Mg system from 16.0
MeV to 24.0 MeV in the laboratory system by using the standard and
new coupled-channels calculations. In the introduction, we
presented the problems that this reaction manifests. We attempted
to find a consistent solution to these problems. However, within
the standard coupled-channels method, we failed, as others did, to
describe certain aspects of the data, in particular, the magnitude
of the 2$^{+}$ excitation inelastic scattering data although the
optical model and coupled-channels models explain perfectly some
aspects of the elastic scattering data. We were compelled to
increase the value of Coulomb deformation to reproduce the 2$^{+}$
data and such arbitrary uses of $\beta$ have been practiced in the
past without giving any physical justifications other that stating
it is required to fit the experimental data.

We have obtained excellent agrement with the experimental data
over the whole energy range by using a new coupling potential,
which has been outstandingly successful in explaining the
experimental data for the the $^{12}$C+$^{12}$C~\cite{Boz1} and
$^{16}$O+$^{28}$Si~\cite{Boz4} systems over wide energies. The
comparison of the results indicates that a global solution to the
problems relating to the scattering observables of this reaction
over a wide energy range has been provided by this new coupling
potential. This work reveals that there is no reason for the
coupling potential to have the same energy dependence as the
central term. The work in order to derive the coupling potential
explicitly from a microscopic viewpoint is still under progress
and studies using this new coupling potential may lead to new
insights into the formalism and also a new interpretation of such
reactions.
\section{Acknowledgments}
Authors wish to thank Doctors Y. Nedjadi, S. Ait-Tahar, R.
Mackintosh, B. Buck, A. M. Merchant, Professor B. R. Fulton and
Ay\c{s}e Odman for valuable discussions and encouragements. I.
Boztosun also would like to thank the Turkish Council of Higher
Education (Y\"{O}K) and Erciyes University, Turkey, for their
financial support.
\begin{table}[h]
\begin{center}
\begin{tabular}{lllllll}
V & $R_{V}$ & $a_{V}$ & $W$ & $R_{W}$ & $a_{W}$ & $R_{c}$\\
(MeV) & (fm) & (fm) & (MeV) & (fm) & (fm) & (fm)\\     \hline
427.0 & 4.486 & 1.187 & Eq. (\ref{cmgimag}) & 1.386 & 0.286 & 5.174\\
\end{tabular}
\end{center}
\caption{The parameters of the real and imaginary potentials. $V$
and $W$ stand for the strengths of the real and imaginary parts
respectively. $R_{c}$ is the Coulomb radius.} \label{realparam}
\end{table}
\begin{table}
\begin{center}
\begin{tabular}{lll}
$E_{Lab}$ & Standard CC &   New CC  \\     \hline
16.0 & 1.24 &   1.4 \\
17.0 & 1.37 &   0.59 \\
18.5 &  0.67 &  0.34 \\
19.0 &  1.44 &  0.8 \\
19.5 &  1.29 &  0.7 \\
20.0 &  1.63 &  0.4 \\
20.51 & 2.48 &  0.5 \\
21.0 & 2.45 &   1.6 \\
21.5 &  3.46 &  3.1 \\
22.0 &  8.2 &   2.3 \\
22.5 &  1.7 &   3.0 \\
23.0 &  3.96 &  1.4 \\
23.5 &  5.69 &  3.5 \\
24.0 &  6.59 &  4.8 \\
\end{tabular}
\end{center}
\caption{The numerical values of $\chi^{2}$ for the standard and
new coupled-channels calculations.} \label{chi1}
\end{table}
\end{document}